\documentclass[a4,letterpape,11pt]{article}

\usepackage{jheppub}

\usepackage{amsmath}
\usepackage{bm}
\usepackage{slashed}
\usepackage{amssymb}
\usepackage{mathtools}
\usepackage{graphicx}
\usepackage{subfigure}
\usepackage{setspace}
\usepackage{sidecap}
\usepackage{color}
\usepackage[colorlinks=true,linkcolor=blue,citecolor=blue]{hyperref}
\usepackage{tabu} 
\usepackage[export]{adjustbox}
\usepackage{tensor}
\usepackage{dcolumn}
\usepackage[normalem]{ulem} 

\newcommand{\lb}{\Big{\lbrack}}
\newcommand{\rb}{\Big{\rbrack}}
\newcommand{\lp}{\Big{(}}
\newcommand{\rp}{\Big{)}}
\newcommand{\lbc}{\Big{\lbrace}}
\newcommand{\rbc}{\Big{\rbrace}}
\newcommand{\nn}{\nonumber}

\newcommand{\bmat}[1]{\boldsymbol{#1}}

\newcommand{\Q}[4]{ {}^{#1} #2 ^{[#4]}_{#3} }

\newcommand{\ot}{\leftarrow}
\newcommand{\jet}{\text{jet}}
\newcommand{\UV}{\text{UV}}
\newcommand{\IR}{\text{IR}}

\newcommand{\vecb}[1]{\mbox{\boldmath $#1$}}

\newcommand{\sandwich}[3]{\left< #1 \right | #2 \left | #3 \right >}
\newcommand{\bn}{{\bar n}}

\renewcommand{\(}{\left(}
\renewcommand{\)}{\right)}
\renewcommand{\[}{\left[}
\renewcommand{\]}{\right]}
\renewcommand{\vec}[1]{\bm{#1}}

\def\t{\tau}
\def \le { \left    }
\def \ri { \right }

\title{Quarkonium TMD fragmentation functions in NRQCD}
\date{\today}
\author[a]{Miguel G. Echevarria,}
\author[b]{Yiannis Makris,}
\author[c]{and Ignazio Scimemi}
\affiliation[a]{Dpto. de F\'{i}sica y Matematicas, Universidad de Alcal\'{a}, 28805 Alcal\'{a} de Henares (Madrid), Spain}
\affiliation[b]{INFN Sezione di Pavia, via Bassi 6, I-27100 Pavia, Italy }
\affiliation[c]{Dpto. de F\'{i}sica Te\'{o}rica \& IPARCOS, Universidad Complutense de Madrid, E-28040 Madrid, Spain}

\emailAdd{m.garciae@uah.es}
\emailAdd{yiannis.makris@pv.infn.it}
\emailAdd{ignazios@ucm.es}

\abstract{
We study the transverse-momentum spectrum of quarkonium production from single light-parton fragmentation mechanism. 
In the case of semi-inclusive deep inelastic scattering, we observe that there are two possible initiating processes, namely photon-gluon fusion and light-quark fragmentation. 
For the second case we derive the factorization theorem, which involves a new hadronic quantity: the quarkonium transverse-momentum-dependent fragmentation functions in NRQCD.
We calculate their matching onto the non-perturbative long distance matrix elements at the lowest order in the strong-coupling constant, ${\mathcal O}(\alpha_s^2)$. 
Focusing on the case of the electron-ion collider, we make a comparative phenomenological study of the two production mechanisms and find the regions of the phase space where one is dominant over the other.
}

\begin{document}

\maketitle

\section{Introduction}
\label{sec:intro}

In recent years, the role of quarkonium production in the study of transverse-momentum-dependent distributions (TMDs) \cite{Angeles-Martinez:2015sea} has been gaining  significant attention. 
Particularly, various processes have been proposed as a probe of gluon TMD parton distribution functions (TMDPDFs) \cite{Echevarria:2015uaa,Mulders:2000sh} in both hadron-hadron and lepton-hadron colliders~\cite{Boer:2012bt,Ma:2012hh,Zhang:2014vmh,Ma:2015vpt, Boer:2015uqa,Bain:2016rrv,Mukherjee:2015smo,Mukherjee:2016cjw,Lansberg:2017tlc,Lansberg:2017dzg,Bacchetta:2018ivt,DAlesio:2019qpk,Echevarria:2019ynx,Fleming:2019pzj,Scarpa:2019fol,Grewal:2020hoc,Boer:2020bbd}. 
In the vast majority of these studies the approach to quarkonium production relies on effective field theories (EFTs) such as the non-relativistic QCD (NRQCD)~\cite{Bodwin:1994jh} or models such as the color evaporation model~\cite{Fritzsch:1977ay, HALZEN1977105, Gluck:1977zm}. 

\begin{figure}
\begin{center}
\includegraphics[width=\textwidth]{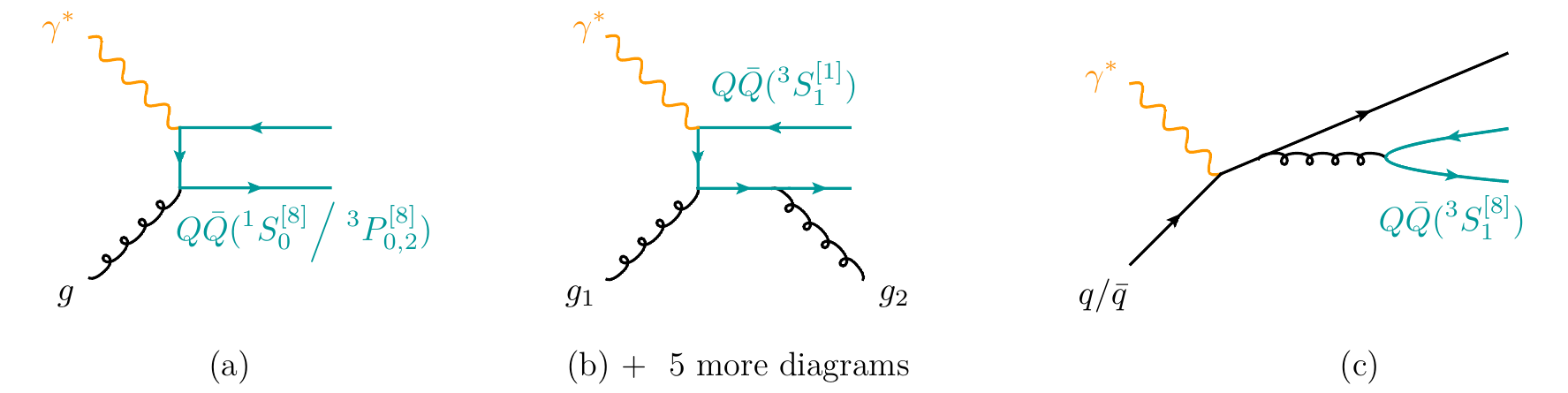}
\caption{\label{fig:B_fig3} Quarkonia generation by double parton fragmentation at leading order (LO) (a) and next-to-leading order (NLO) (b). 
In (c) we have the quarkonia generation by single-quark fragmentation at LO.}
\end{center}
\end{figure}

Despite the abundant interest in quarkonium TMDs, little has been done in the direction of TMD quarkonium fragmentation processes, by which we mean single parton fragmentation mechanism (see for example figure~\ref{fig:B_fig3}(c)). 
The fragmentation process becomes relevant when a hard scale, $\Lambda$, much larger than the quarkonium mass, $M$, exists. 
For example, in hadronic colliders the quarkonium TMD fragmentation can be accessed inside jets where quarkonia are found with relatively large transverse momenta. 
In this scenario,  the hard scale of the problem is set by the transverse momentum of the jet,  $\Lambda = p_T^{\jet}$, and the TMD spectrum of quarkonium is then measured w.r.t. the  jet axis.  
This type of studies can provide important insight into the heavy-quark hadronization mechanisms and, in terms of contamination from the underlying event (UE), the study of quarkonia is considered advantageous compared to light hadrons.~\footnote{The effect of soft background radiation in a jet can be further minimized using modern jet substructure techniques such as grooming or recoil free jet axis. It is important to note that for such jet modifications theoretical calculations are possible, thus permitting rigorous and precise comparisons of theory with experiment. For  TMD studies of in-jet light hadrons using such techniques see refs.~\cite{Neill:2016vbi, Makris:2017arq, Neill:2018wtk} and for heavy mesons see ref.~\cite{Makris:2018npl}.} 

On the other hand, in semi-inclusive deep inelastic scattering (SIDIS) the hard scale $\Lambda$ is usually set by the invariant mass of the virtual photon, $\Lambda=Q = \sqrt{-q^2}$, with $q$ the photon momentum. 
Thus, performing the TMD measurement with respect to the photon's direction in the Breit frame we can formulate a global TMD factorization theorem in a similar manner to light hadrons. 
Since in an electron-hadron collider the kinematical range of $x$ and $Q^2$ is limited, in the present study we propose to explore the window in which single-quarkonium TMD fragmentation is relevant compared to the photon-gluon fusion quarkonium production.

In our analysis we employ the NRQCD factorization conjecture where quarkonium is produced at large distances through the hadronization of a heavy quark-antiquark pair, $Q\bar{Q}(n)$. The pair can be found in any color and angular configuration $n=\Q{2S+1}{L}{J}{\text{col.}}$ but then the probability that the pair decays in the colorless quarkonium state scales with the relative velocity, $v$, of the quark-antiquark pair in the quarkonium rest frame. In this study we are primarily interested in the quarkonium state $J/\psi$, for which the four leading channels in $v$ are: $\Q{3}{S}{1}{1}(\sim v^3)$, $\Q{3}{S}{1}{8}(\sim v^7)$, $\Q{1}{S}{0}{8}(\sim v^7)$, and $\Q{3}{P}{J}{8}(\sim v^7)$. 

We  are going to make our case for the forthcoming electron-ion collider (EIC). 
In order to start our discussion let us explore quarkonium production for this type of experiment. 
At leading order (LO) in the perturbative expansion, quarkonium production is possible through the photon-gluon fusion process as shown in figure~\ref{fig:B_fig3}(a). 
In this case the heavy-quark pair is produced either in $\Q{1}{S}{0}{8}$ or $\Q{3}{P}{J=0,2}{8}$ states, the transverse momentum of the heavy-quark pair vanishes and $z\equiv p_h\cdot P/p_h\cdot q =1$ (where $p_h$ is the target momentum and $P$ the heavy-quarkonium momentum).  
At higher orders a smearing around the Born kinematics will occur due the emissions of soft and collinear gluons from the incoming partons as well as the heavy quark-antiquark state. 
The cross section for these processes can be organized in terms of the TMDPDFs and the recently introduced TMD quarkonium shape-functions~\cite{Echevarria:2019ynx, Fleming:2019pzj}. 
At $\mathcal{O}(\alpha_s^2)$ we also have a contribution from the $\Q{3}{S}{1}{1}$ (see figure~\ref{fig:B_fig3}(b)). 
Although this contribution is power suppressed in $\bmat{q}^2_T /Q^2$, enhancements from the relative velocity scaling make it hard to argue that this channel can be neglected. 
At the same order in $\alpha_s$ enters the quarkonium fragmentation process shown in figure~\ref{fig:B_fig3}(c), whose relative channel is then $\Q{3}{S}{1}{8}$.

To isolate the fragmentation process we take advantage of the fact that both $\Q{3}{S}{1}{8}$ and $\Q{3}{S}{1}{1}$ at Born level have continuous distributions in $z$ and, thus, focussing on the region $z \lesssim 0.5$ allows us to better isolate these two channels. 
In addition, to further suppress the non-fragmentation contributions we work in the region $Q \gg M \sim \bmat P_{\perp}$ (for details see discussion in section~\ref{sec:gluon-photon}). 
We explore here how the competition of gluon-photon fusion and quark fragmentation depends on the kinematical conditions in which an experiment is run, in order to establish the study case of this process.

The factorization of the TMD cross section for the light-quark fragmentation mechanism is structurally very similar to the conventional SIDIS process~\cite{Collins:2011zzd,GarciaEchevarria:2011rb,Echevarria:2012js,Chiu:2012ir}, where the rapidity scale $\zeta$ and the renormalization scale $\mu$ are responsible of the evolution of the TMD matrix elements in initial and final states~\cite{Echevarria:2012pw}. 
This process has been extensively studied in the literature and we now have high-order QCD calculations for hard factors~\cite{Kramer:1986sg,Matsuura:1988sm,Gehrmann:2010ue}, evolution kernel~\cite{Echevarria:2015byo,Vladimirov:2016dll,Li:2016ctv} and TMDPDFs~\cite{Gehrmann:2014yya,Luebbert:2016itl,Echevarria:2015usa,Echevarria:2016scs,Luo:2019hmp,Luo:2019szz, Ebert:2020yqt}. 
It is also possible then to use extraction of TMDs~\cite{Bacchetta:2017gcc, Scimemi:2019cmh, Bacchetta:2019sam} to perform a phenomenological analysis. 
In this work we use the extractions done in ref.~\cite{Scimemi:2019cmh} with the so-called $\zeta$-prescription~\cite{Scimemi:2018xaf} and the associated Artemide code~\cite{web}. 
 
The quarkonium TMD fragmentation function (TMDFF), $D_{ f\to H}$, is the only piece which remains to be studied and included in our analysis. 
We decompose the TMDFF within NRQCD in terms of calculable short-distance matching coefficients and long-distance matrix elements (LDMEs). 
We proceed then to the LO calculation of this function, extracting the matching coefficient onto the corresponding LDMEs in the region $q_T \sim M$. 
Using this information we can evaluate the contribution to the cross section from light-quark fragmentation.

This paper is organized as follows: we introduce our notation and the form of the factorization theorem for light quark fragmentation in section~\ref{sec:notation}. 
In section~\ref{sec:quark} we perform the calculation of the matching coefficient of the TMDFF onto the $\Q{3}{S}{1}{8}$ LDME at ${\mathcal O}(\alpha_s^2)$. 
We also include a short discussion on the RG evolution of LDMEs which becomes relevant for TMD calculations due to the additional soft scale $q_T$. 
In section~\ref{sec:gluon-photon} we give an analytic comparison of the fixed-order results for the $\Q{3}{S}{1}{1/8}$ channels for DIS in the small transverse momentum limit. 
In section~\ref{sec:numerics} we demonstrate the phenomenological applicability of this formalism for the future EIC kinematics and provide for the first time numerical estimates of the relative importance of the fragmentation channel in various kinematic regions. 
Finally, in section~\ref{sec:outlook} we conclude.

\section{Notation}
\label{sec:notation}

The momenta  of quarkonium production in SIDIS are specified by 
\begin{eqnarray}
\label{eq:s1}
\ell(l)+h(p)\to \ell(l')+H(P)+X,
\end{eqnarray}
where $\ell$ is a lepton, $h$ and $H$ are respectively the initial and the final hadrons, and $X$ is the undetected final state. The masses of the hadrons are
\begin{eqnarray}
p^2=m^2,\qquad P^2=M^2,
\end{eqnarray}
and we neglect lepton masses. For the moment we continue to consider hadron masses in order to control their effects.

The differential cross-section in eq.~(\ref{eq:s1}) can be written as 
\begin{eqnarray}\label{def:xSec-0}
d\sigma=\frac{2}{s-m^2} \frac{ \alpha^2_{\text{em}}}{(q^2)^2} L_{\mu\nu}W^{\mu\nu}\frac{d^3l'}{2E'}\frac{d^3P}{2E_H},
\end{eqnarray}
with $q=l-l'$ being the momentum of the intermediate photon and  $\alpha_{\text{em}}=e^2/4\pi$ the QED coupling constant. 
The cross section in eq.~(\ref{def:xSec-0}) is proportional to the phase-space differentials for the detected lepton and heavy hadron, with $E'$ and $E_H$ being their energies. The leptonic and hadronic tensors ($L^{\mu\nu}$ and $W^{\mu\nu}$) are \begin{eqnarray}
L_{\mu\nu}&=&e^{-2}\langle l'|J_{\mu}(0)|l\rangle \langle l|J^\dagger_{\nu}(0)|l'\rangle,\nn
\\
W_{\mu\nu}&=&e^{-2}\int \frac{d^4x}{(2\pi)^4}e^{-i(x q)}\sum_X\langle p|J^\dagger_{\mu}(x)|P,X\rangle \langle P,X|J_{\nu}(0)|p\rangle,
\end{eqnarray}
where $e$ is the lepton charge, and $J_\mu$ is the electro-magnetic current.

\subsection{Kinematical variables for SIDIS}

In SIDIS one makes use of different frames which specify the hadronic variables. In this section we follow closely \cite{Scimemi:2019cmh}, adapting the variables to the heavy quarkonium case.
Although in our numerical evaluation we  only consider some approximation of the  kinematical variables, we prefer to write them down for future reference.
The factorization theorem for the cross section is done in the Breit frame, where the momenta of hadrons are back to back and they are respectively
\begin{eqnarray}
p^\mu&=&p^+ \bar n^\mu+\frac{m^2}{2p^+}n^\mu,\qquad P^\mu =P^- n^\mu+\frac{M^2}{2P^-}\bar n^\mu,
\end{eqnarray}
with
$n^2=\bar n^2=0,$ $ (n\bar n)=1$, and the   vector decomposition
\begin{eqnarray}
v^\mu=v^+ \bar n^\mu+v^- n^\mu+v_T^\mu,\qquad v^+=(nv),\qquad v^-=(\bar nv),\qquad (nv_T)=(\bar nv_T)=0.
\end{eqnarray}
The transverse component of a vector is defined by the projection
\begin{eqnarray}\label{def:gT}
v_T^\mu=g_T^{\mu\nu}v_\nu,\qquad g_T^{\mu\nu}=g^{\mu\nu}-n^\mu \bar n^\nu-\bar n^\mu n^\nu.
\end{eqnarray}
We also use the convention that the bold font denotes vectors that have only transverse components
and 
\begin{eqnarray}
\vec v_T^2=-v_T^2>0.
\end{eqnarray}

The kinematical scalar variables are defined as:
\begin{eqnarray}
Q^2=-q^2,\qquad x=\frac{Q^2}{2(pq)},\qquad y=\frac{(pq)}{(pl)},\qquad z=\frac{(pP)}{(pq)}.
\end{eqnarray}
Experimentally one needs to define a plane transverse to $q$ and $p$, and the projector
 corresponding  to the components  transverse to this plane   is given by the tensor $g_\perp^{\mu\nu}$ defined as
\begin{eqnarray}\label{def:gPerp-SIDIS}
g_\perp^{\mu\nu}&=&g^{\mu\nu}-\frac{1}{m^2Q^2+(pq)^2}\[Q^2 p^\mu p^\nu+(pq)(p^\mu q^\nu+q^\mu p^\nu)-m^2 q^\mu q^\nu\]
\\\nn
&=&g^{\mu\nu}-\frac{1}{Q^2(1+\gamma^2)}\[2x^2 p^\mu p^\nu+2x(p^\mu q^\nu+q^\mu p^\nu)-\gamma^2 q^\mu q^\nu\].
\end{eqnarray}
Using this notation we can distinguish the transverse components of $v^\mu$ in the hadronic Breit frame as $v_T^\mu$, see eq.~(\ref{def:gT}), from the  transverse components projected by $g_\perp$, that is $v_\perp^\mu$. 

The mass corrections are conveniently   described using the following combinations
\begin{eqnarray}\label{def:mass-var}
\gamma=\frac{2mx}{Q},\qquad \varsigma=\frac{\gamma}{z}\frac{M}{Q},\qquad \varsigma^2_\perp=\frac{\gamma^2}{z^2}\frac{M^2+\vec P^2_{\perp}}{Q^2}.
\end{eqnarray}
The definition of $\varsigma^2_\perp$ in eq.~(\ref{def:mass-var}) contains ${\vec P}^2_{\perp}=-P_{\mu} P_{ \nu } g_\perp^{\mu\nu}= -P^2_{\perp}$. 
Using these variables we can re-express
$g_T^{\mu\nu}$ in eq.~(\ref{def:gT}) in terms of hadronic momenta
\begin{eqnarray}\label{def:gT-SIDIS}
g_T^{\mu\nu}&=&g^{\mu\nu}-\frac{1}{m^2M^2-(Pp)^2}\[M^2 p^\mu p^\nu-(Pp)(P^\mu p^\nu+p^\mu P^\nu)+M^2 P^\mu P^\nu\]
\\\nn
&=&g^{\mu\nu}+\frac{1}{Q^2(1-\varsigma^2)}\[4\frac{x^2}{\gamma^2}\varsigma^2 p^\mu p^\nu-\frac{2x}{z}(P^\mu p^\nu+p^\mu P^\nu)+\frac{\gamma^2}{z^2} P^\mu P^\nu\].
\end{eqnarray}
With  eq.~(\ref{def:gPerp-SIDIS}) and eq.~(\ref{def:gT-SIDIS}) one can derive the relation between $q_T^2=q_\mu q_\nu g_T^{\mu\nu}$ and $P_\perp^2$,
\begin{eqnarray}\label{def:qT<->pT}
q_T^2=\frac{P_\perp^2}{z^2}\frac{1+\gamma^2}{1-\varsigma^2}
\,,
\end{eqnarray}
including mass corrections.

Using these definition we can rewrite the elements of the SIDIS cross-section formula in terms of experimental variables, that is,   the differential volumes of the phase space are
\begin{eqnarray}\label{th:phase-elem-1}
\frac{d^3l'}{2E'}=\frac{y}{4x}dQ^2dx d\psi,\qquad \frac{d^3P}{2E_H}=\frac{1}{\sqrt{1-\varsigma^2_\perp}}\frac{dz d^2P_\perp}{2z}
=\frac{1}{\sqrt{1-\varsigma^2_\perp }}\frac{dz d\vec P^2_\perp d \varphi}{4z},
\end{eqnarray}
where $\psi$ is the azimuthal angle of scattered lepton, and $\varphi$ is the azimuthal angle of the produced hadron (we recall that we are in the Breit frame where the $z$-axis is in the proton-photon direction). 

Finally we  introduce the variables $x_S$ and $z_S$, that are the collinear fractions of parton momentum that include kinematic power corrections,
\begin{eqnarray}\label{def:x1z1}
x_S=-\frac{q^+}{p^+},\qquad z_S=\frac{P^-}{q^-},
\end{eqnarray}
which are invariant under boosts along the directions $n$ and $\bar n$, but are not invariant under a generic Lorentz transformation.
The variables $x_S$ and $z_S$ in eq.~(\ref{def:x1z1}) are
\begin{eqnarray}\label{def:SIDIS-x1}
x_S&=&-x\frac{2}{\gamma^2}\(1-\sqrt{1+\gamma^2\(1-\frac{\vec q_T^2}{Q^2}\)}\),
\\\label{def:SIDIS-z1}
z_S&=&-z\frac{1-\sqrt{1+\gamma^2\(1-\frac{\vec q_T^2}{Q^2}\)}}{\gamma^2}\frac{1+\sqrt{1-\varsigma^2}}{1-\frac{\vec q_T^2}{Q^2}}=z\frac{x_S}{x}\frac{1+\sqrt{1-\varsigma^2}}{2\(1-\frac{\vec q_T^2}{Q^2}\)}.
\end{eqnarray}

The kinematic corrections presented above are usually small when  $Q\gg M,m$. In this case the relation between observed and factorization variables simplifies:
\begin{eqnarray}
\label{eq:APPX}
\vec q_T^2\simeq \frac{\vec P_\perp^2}{z^2},\qquad x_S\simeq x,\qquad z_S\simeq z.
\end{eqnarray}
The next correction to this limit happens when $Q\gg M\sim q_T\gg m$, that is in the limit $\gamma\rightarrow 0$ with $\varsigma\sim {\cal O}(1)$.
In this limit
\begin{eqnarray}
q_T^2\simeq \frac{P_\perp^2}{z^2}\frac{1}{1-\varsigma^2}\,,\qquad 
x_S\simeq x \left(1-\frac{\vec q_T^2}{Q^2}\right)\,,\qquad  
z_S\simeq z \frac{1+\sqrt{1-\varsigma^2}}{2}.
\end{eqnarray}
Notice that  in this case there is a shift  effect in the variable $z_S$ which is independent of the transverse momentum. 
Nevertheless, the effect of hadron masses can be considered as a power correction to the process whose impact can be estimated numerically.

\subsection{Factorization of the hadronic tensor }

For $Q \gg M$ the proof of the factorization theorem for quarkonium fragmentation follows the same steps as for the light-hadron case, so we refer the reader to the literature~\cite{Collins:1981va,Collins:1984kg,Collins:1989gx,Catani:1996yz,Collins:2011zzd,GarciaEchevarria:2011rb,Echevarria:2012js,Catani:2013tia} for a detailed treatment.
The factorized hadronic tensor reads
\begin{align}
W^{\mu\nu}=-\frac{ z_S}{\pi} g^{\mu\nu}_T \sum_f e_f^2  \Theta_f(Q,|{\vec q_T}|,x_S,z_S)+O\(\frac{q_T^2}{Q^2}, \frac{M^2}{Q^2}\),
\end{align}
where
\begin{eqnarray}\label{SIDIS:Wmunu}\nn
\Theta_f(Q,|{\vec q_T}|,x_S,z_S)&=&2 \pi |C_V(Q^2,\mu^2)|^2 \int \frac{d^2\vec b}{(2\pi)^2}e^{-i(\vec b\vec q_T)} 
f_{1,f\ot h}\(x_{S},b;\mu,\zeta_1\)D_{f\to H}\(z_S,b;\mu,\zeta_2\)
\\\nn && 
+O\(\frac{q_T^2}{Q^2}, \frac{M^2}{Q^2}\),
\\\nn &=& |C_V(Q^2,\mu^2)|^2 \int_0^\infty bdb\; J_0(b |\vec q_T|)f_{1,f\ot h}\(x_{S},b;\mu,\zeta_1\)D_{f\to H}\(z_S,b;\mu,\zeta_2\)
\\ &&
+O\(\frac{q_T^2}{Q^2}, \frac{M^2}{Q^2}\)
\,.
\end{eqnarray}
$J_0$ is the Bessel function, the index $f$ runs through all quark flavors (including anti-quarks) and $e_f$ is the fractional charge of a quark measured in units of $e$. 
The function $C_V$ is the matching coefficient for the vector current.
We also especify the factorization ($\mu$) and rapidity ($\zeta$) scales, and we have omitted the details of operator definitions, such as $T$($\bar T$)-ordering, color and spinor indices, and rapidity and ultraviolet renormalization factors.
 
The bare unsubtracted unpolarized TMDPDF and quarkonium TMDFF are defined as
\begin{eqnarray}\label{def:f1}
&&\Phi_{1,q\ot h}(x,b)=
\\\nn &&\qquad\int \frac{d\lambda}{2\pi}e^{-ix\lambda p^+}
\sum_X \langle h(p)|\bar q(n \lambda+b) W^\dagger_n(n\lambda+b)\frac{\gamma^+}{2}|X\rangle\langle X|W_n(0)q(0)|h(p))\rangle,
\\\label{def:D1}
&&\Delta_{q\to H}(z,b)=
\\\nn &&\qquad\frac{1}{2zN_c}\int \frac{d\lambda}{2\pi}e^{i\lambda P^-/z}\sum_X\langle 0|\frac{\gamma^-}{2}W_{\bar n}(\bar n\lambda+b) q(\bar n \lambda+b)|H(P),X\rangle\langle H(P),X|\bar  q(0)W^\dagger_{\bar n}(0)|0\rangle
\,,
\end{eqnarray}
and similarly for antiquarks.
Here, $W_v(x)$ are Wilson lines at $x$ and pointing along vector $v$ at infinity. 
In the case of SIDIS, the Wilson lines in the TMDPDF (TMDFF) are pointing respectively at past (future) infinity.
Notice that the operator that appears in the fragmentation function is the same as in the case of light quarks and that the only change is the presence of a quarkonium in the final state.
 
The renormalized TMDs are defined in the usual way
\begin{align}
F_{q\ot {h}}(x,b,\mu,\zeta)&=Z_{2}^{-1}(\mu)Z_q(\zeta,\mu)R_q(\zeta,\mu)\Phi_{q\ot h}(x,b),\nn\\
D_{q \to H}(z,b,\mu,\zeta)&=Z_{2}^{-1}(\mu)Z_q(\zeta,\mu)R_q(\zeta,\mu)\Delta_{q\to H}(z,b)\;,
\end{align}
where $Z_{2}$  is the wave function renormalization constant for quarks and $Z_q$ the renormalization factor for the UV divergences of the TMD operator. 
The factor $R_q$ is the ``rapidity renormalization factor''.   Its definition comes from the TMD factorization theorem and reads
\begin{eqnarray}
R_q(\zeta,\mu)=\frac{\sqrt{S(b)}}{\textbf{Zb}},
\end{eqnarray}
where $S(b)$ is the soft function and \textbf{Zb} denotes the zero-bin contribution, i.e.  the soft overlap of the collinear and soft sectors  present in the factorization theorem~\cite{Manohar:2006nz,Collins:2011zzd,GarciaEchevarria:2011rb,Echevarria:2012js}.

The soft function  in the present case  is the same as in the light quark case;  for SIDIS it reads (here $b=|\vecb b_T|$)
\begin{eqnarray} \label{eq:SF_def}
S(b)
=
\frac{{\rm Tr}_c}{N_c} \sandwich{0}{\, T\le[S_n^{T\dagger} \tilde S_\bn^T \ri](0^+,0^-,\vecb b_T) \bar
T\le[\tilde S^{T\dagger}_\bn S_n^T\ri](0)}{0}\,.
\end{eqnarray}
The Wilson lines are defined as usual
\begin{eqnarray}
 \label{eq:SF_def2}
S_{n}^T &=& T_{n} S_{n}\,,
\quad\quad\quad\quad
\tilde S_{\bn}^T = \tilde T_{n} \tilde S_{\bn}\,,
\\ \nn
S_n (x) &=& P \exp \left[i g \int_{-\infty}^0 ds\, n \cdot A (x+s n)\right]\,,
\\\nn
T_{\bn} (x) &=& P \exp \left[i g \int_{-\infty}^0 d\tau\, \vec l_\perp \cdot \vec A_{\perp} (0^+,\infty^-,\vec x_\perp+\vec l_\perp \tau)\right]\,,
\\\nn
\tilde S_\bn (x) &=& P\exp\le[-ig\int_{0}^{\infty} ds\, \bn \cdot A(x+\bn s) \ri]\,,
\\\nn
\tilde T_{n} (x) &=& P\exp\le[-ig\int_{0}^{\infty} d\t\, \vec l_\perp \cdot \vec A_{\perp}(\infty^+,0^-,\vec x_\perp+\vec l_\perp\t) \ri]\,.
\nn
\end{eqnarray}
The transverse gauge links $T_{n(\bn)}$ are to be used in singular gauges, like the light-cone gauge $n \cdot A=0$ (or $\bn \cdot A=0$)~\cite{Belitsky:2002sm,Idilbi:2010im,GarciaEchevarria:2011md}.

The definition of the zero-bin depends on the rapidity regularization used in loop calculations (see e.g. discussion in \cite{Echevarria:2012js}).
With the modified $\delta$-regularization, the zero-bin subtraction is  equal to the soft function: $\textbf{Zb}=S(b)$.
As a result in the modified $\delta$-regularization   the rapidity renormalization factor is
\begin{eqnarray}\label{reg:R=1/S}
R_q(\zeta,\mu)\bigg|_{\delta\text{-reg.}}=\frac{1}{\sqrt{S(\vecb{b}_T;\zeta)}}.
\end{eqnarray}
Due to the process independence of the soft function \cite{Echevarria:2014rua}, the factor $R_q$ is also
process independent.

 \subsection{Leptonic tensor and cross section}

The leptonic tensor for our case is the same as for the unpolarized SIDIS process:
\begin{eqnarray}
L_{\mu\nu}=2(l_\mu l'_\nu+l'_\mu l_\nu- (ll') g_{\mu\nu} ).
\end{eqnarray}
The contraction of the leptonic tensor with the hadronic tensor can be performed with the definition of  the azimuthal angle of a produced hadron as \cite{Bacchetta:2006tn}:
\begin{eqnarray}\label{th:SIDIS-L1}
(-g_T^{\mu\nu}) L_{\mu\nu}&=&\frac{2Q^2}{1-\varepsilon}\Big[1+\frac{\vec P_\perp^2}{Q^2 z^2}\frac{\varepsilon-\frac{\gamma^2}{2}}{1-\varsigma^2}
\\\nn &&
 -\cos\phi \frac{\sqrt{2\varepsilon(1+\varepsilon)\vec P_\perp^2}}{zQ}\frac{\sqrt{1-\varsigma_\perp^2}}{1-\varsigma^2}
 -\cos(2\phi)\frac{\varepsilon\vec P_\perp^2 \gamma^2 }{2 z^2 Q^2(1-\varsigma^2)}
\Big]
\,,
\end{eqnarray}
where
\begin{eqnarray}
\cos\phi=\frac{-l_\mu p_{h\nu}g_\perp^{\mu\nu}}{\sqrt{-l_\alpha l_\beta g_\perp^{\alpha\beta} }\sqrt{-p_{h\alpha'} p_{h\beta'} g_\perp^{\alpha'\beta'} }}
\end{eqnarray}
and
$$
\varepsilon=\frac{1-y-\frac{\gamma^2y^2}{4}}{1-y+\frac{y^2}{2}+\frac{\gamma^2 y^2}{4}}
\,.
$$
We observe that $\phi=\varphi$ when the $x$-axis is chosen to lay on the lepton plane along to the outgoing lepton direction, which is the usual choice \cite{Bacchetta:2006tn}.
As in the light-quark case, the kinematical rearrangements of the variables contribute to the  $\cos\phi$ and $\cos2\phi$ terms in the second line of eq.~(\ref{th:SIDIS-L1}),
 which disappear when one integrates over angles~\cite{Anselmino:2005nn}.

Recollecting the equations for the cross-section in eq.~(\ref{def:xSec-0}), the differential phase-space volume in eq.~(\ref{th:phase-elem-1}), the hadronic tensor in eq.~(\ref{SIDIS:Wmunu}), the leptonic tensor in eq.~(\ref{th:SIDIS-L1}),  and integrating over the azimuthal angles  we get
\begin{eqnarray}\label{SIDIS:xSec}
&&\frac{d\sigma}{dxdz dQ^2 d\vec P_\perp^2}=\frac{\pi}{\sqrt{1-\varsigma_\perp^2}}\frac{\alpha_{\text{em}}^2}{Q^4}\frac{y^2}{1-\varepsilon}\frac{z_S}{z}
\sum_{f}e_f^2
\(1+\frac{\vec q_T^2}{Q^2}\frac{\varepsilon-\frac{\gamma^2}{2}}{1+\gamma^2}\)\Theta^f(Q,|\vec q_T|,x_S,z_S)\;,
\end{eqnarray}
where $\vec q^2_T$, $x_S$ and $z_S$ are functions of $\vec P_\perp^2$, $x$ and $z$, and are defined respectively in eqs.~(\ref{def:qT<->pT}), (\ref{def:SIDIS-x1}) and~(\ref{def:SIDIS-z1}).
In eq.~(\ref{SIDIS:xSec}) we have reported all calculable power corrections which  come from the kinematics of the process. 
Other power corrections of the same order can be obtained from the expansion of the cross section. 
Since their study goes beyond the purpose of this work and given that they are still largely unexplored, we omit them in the present study.

\section{Quarkonium TMDFF from a light quark}
\label{sec:quark}

In the present section we report the calculation of  the  TMD fragmentation function  as defined in eq.~(\ref{def:D1}). 
In our approach and for $q_T \sim M$, we employ NRQCD factorization conjecture~\cite{Bodwin:1994jh} to write the quarkonium TMDFF as a product of short distance coefficients, $d_{q \to Q\bar{Q}(n)}$, and the standard NRQCD long distance matrix elements (LDMEs):
\begin{equation}
\label{eq: nrqcd-match}
D_{q \to H} (z,b;\mu,\zeta) = \sum_n d_{q \to Q\bar{Q}(n)} (z,b;\mu,\zeta) \frac{ \langle \mathcal{O}^{H}(n) \rangle}{N_{\text{col.}} N_{\text{pol.}}} \;.
\end{equation}
where $N_{\text{col.}}$ and $N_{\text{pol.}}$ are the number of colors and polarizations, respectively, of the heavy quark pair $Q\bar{Q}(n)$. 
The short-distance coefficients can be calculated perturbatively through matching. 
Since they do not depend on the quarkonium state $H$, they can be calculated perturbatively by choosing an appropriate partonic state. 
On the other hand, the LDMEs $\langle \mathcal{O}^{H}(n) \rangle$ are non-perturbative numbers extracted from experimental studies, which scale with the relative velocity, $v$, of the heavy quark-antiquark pair in the quarkonium rest frame.

\begin{figure}
\includegraphics[width=\textwidth]{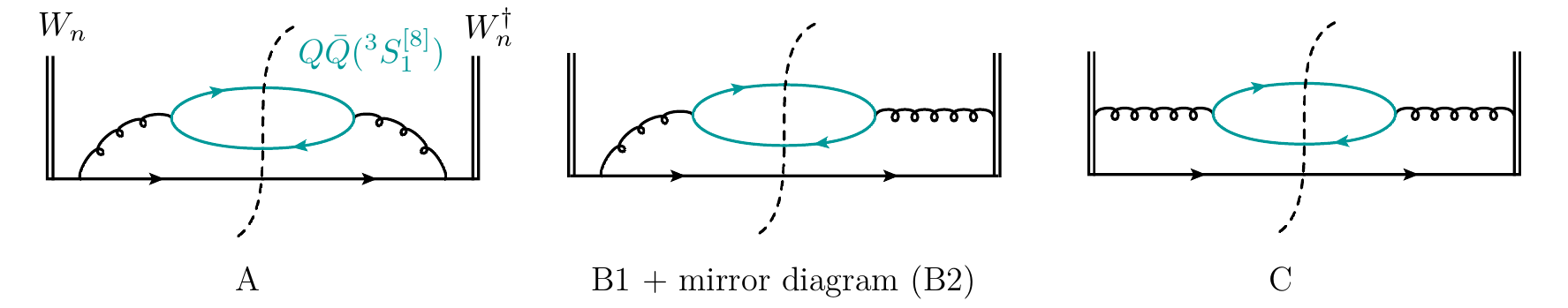}
\caption{\label{fig:B_fig2} Lowest order diagrams for quarkonium TMDFF from a light quark.}
\end{figure}

In this study we are primarily interested in the quarkonium state $H=J/\psi$ for which the four leading channels  in $v$ are: $\Q{3}{S}{1}{1}(\sim v^3)$, $\Q{3}{S}{1}{8}(\sim v^7)$, $\Q{1}{S}{0}{8}(\sim v^7)$ and $\Q{3}{P}{J}{8}(\sim v^7)$. 
However, at leading order (LO) in the strong coupling only the $n =\Q{3}{S}{1}{8}$
intermediate state contributes and, thus, in $d$ space-time dimensions we can write 
\begin{equation}
\label{eq:NRQCD}
D_{q \to \psi} (z,b;\mu,\zeta) =  d_{q \to Q\bar{Q}(\Q{3}{S}{1}{8})} (z,b;\mu,\zeta) \frac{\langle \mathcal{O}^{\psi}(\Q{3}{S}{1}{8}) \rangle}{(d-1)(N_c^2 - 1)} \lp 1  + O (\alpha_s) \rp
\,,
\end{equation}
where we use the shorthand notation $\psi \equiv J/\psi$.  
Contributions to light-quark fragmentation at LO in pQCD are shown in figure~\ref{fig:B_fig2}. 
Note that in our convention here the short distance matching coefficients are defined after renormalization and the inclusion of the soft functions and zero-bin subtraction. 
However, at the order we are working the soft function contribution is trivial and therefore we have only contributions from the direct matching of the collinear matrix element onto the NRQCD LDME. 
This is consistent with the fact that there are no rapidity divergent diagrams at this order. 
These will occur at higher orders in the perturbative expansion, where the soft function will enter into play to cancel them.

In the calculation of the diagrams A, B1, B2, and C  the only relevant master integral that we have to calculate is 
\begin{align}\label{eq:i1}
F_n(B\Delta)=M^{2 n}
\int \frac{d^{2-2\epsilon}k_T}{(2\pi)^{2-2\epsilon}} \frac{e^{i \mathbf{k_T b} }}{(k_T^2+\Delta)^{n+1}}
=\frac{2 M^{2 n}}{(4\pi)^{1-\epsilon} \Gamma[n+1]}
\left(\frac{B}{\Delta }\right)^{\frac{n+\epsilon}{2}} K_{-n-\epsilon}\left(2 \sqrt{B \Delta }\right)
\,,
\end{align}
with $B=b^2/4$ and $K_i$ is the $i$-th Bessel-$K$ function. In the perturbative calculation we used the approximation $m_c \simeq M/2$. The results for each diagram ($d_i$) reported in figure~\ref{fig:B_fig2} are (we use the common notation $\bar z\equiv1-z$ and $\Delta\equiv M^2 \bar z/z^2 $)
 \begin{align}
 d_A&= 4 \pi \alpha_s^2 C_F \frac{I_0}{ M^4  z^3} 
\lb (\epsilon -1) \left(2 \bar z F_1(B\Delta)-2 z^2 F_0(B\Delta)\right)+ z^2 F_{-1}(B\Delta) \rb  \mu^{2 \epsilon}\, ,
 \\
d_{B1+B2}&= 8 \pi \;\alpha_s^2 \;C_F \frac{I_0}{ M^4   z^3}  \lb 2\bar z F_0(B\Delta)- z^2F_{-1}(B\Delta) \rb \mu^{2 \epsilon} \, ,
\\
d_C&= 4 \pi  \alpha_s^2\; C_F \frac{I_0}{ M^4   z}
   F_{-1}(B\Delta)\mu^{2 \epsilon}\,,
\\
\label{eq:res1}
d_A+d_{B1+B2}+d_C&= - 8 \pi \; \alpha_s^2 \;C_F \frac{I_0}{ M^4   z^3} \lb  (1-\epsilon)  \bar{z }F_1(B\Delta) - (  z^2(1-\epsilon)+2\bar z ) F_0(B\Delta) \rb  \mu^{2 \epsilon}
\,,
\end{align}
where we have used
\begin{align}
\frac{1}{M^4}  \sum_{\rm pol.} [\bar u(p)\gamma^\mu T^A v(p')][\bar v(p')\gamma^\mu T^A u(p)] \Big{\vert}_{p-p'=0,\;p+p'=P}
=  & \frac{\Lambda^\mu_i\Lambda^\nu_j}{M^2} \sum_{\rm pol.}[\xi^\dagger  \sigma^iT^A\eta ]
[\eta^\dagger \sigma^j T^A \xi ]   \nn\\ 
= &\frac{(N_c^2-1)}{M^4} \lp -g^{\mu\nu}+\frac{P^\mu P^\nu}{M^2}\rp \; I_0 
\end{align}
with
\begin{align}
I_0= M^2 \frac{ \sum_{\rm pol.}[\xi^\dag \sigma^iT^A \eta]
[\eta^\dag \sigma^i T^A \xi] 
}{(d-1)(N_c^2-1)}
\,.
\end{align}
$\Lambda$'s are the Lorentz boosts from the heavy quarkonium center of mass frame to the boosted collinear frame and the expansion of the spinors is performed using the formulas found in the appendix of ref.~\cite{Braaten:1996rp}. With this we have completed the QCD part of the calculation. 

The pNRQCD part of the calculation is rather simple. We first rewrite the LDME in terms of the relativistically normalized matrix element
\begin{equation}
2M \langle \mathcal{O}^{\psi} (\Q{3}{S}{1}{8}) \rangle = \langle \chi^{\dag} \sigma^{i} T^{A}\psi \; \mathcal{P}_{\psi} \; \psi^{\dag} \sigma^{i} T^{A}\chi  \rangle 
\,,
\end{equation}
where 
\begin{equation}
 \mathcal{P}_{\psi}  =  \sum_{\lambda} \mathcal{P}_{\psi (\lambda)} .\;
\end{equation}
We then perform the perturbative calculation by  replacing  
\begin{equation}
 \sum_{\lambda} \mathcal{P}_{\psi (\lambda)}  \to  \sum_{\text{pol.}} \mathcal{P}_{Q\bar{Q}(^3S^{(8)}_{1})} .\;
\end{equation}
This then yields 
\begin{equation}
2M \langle \Q{3}{S}{1}{8} \rangle \Big{\vert}_{\text{pNRQCD}} =  \sum_{\rm pol.}[\xi^\dag \sigma^iT^A \eta]
[\eta^\dag \sigma^i T^A \xi]  \lp 1+  O (\alpha_s) \rp .
\end{equation}
From these we immediately have (see eq.~(\ref{eq:NRQCD}))
\begin{equation}
\label{eq:TMD-FF-b}
d_{q\to Q\bar{Q}(^3S_1^{8})}(z,b;\mu,\zeta) = -16 \pi \frac{\alpha_s^2 \;C_F}{ M^3 z^3} \lb  (1-\epsilon) \bar{z }F_1(B\Delta) - (  z^2(1-\epsilon)+2\bar z ) F_0(B\Delta) \rb  \mu^{2 \epsilon}\,.
\end{equation} 
We also observe that the most divergent part of diagrams cancel in the sum. It is interesting to check also this  result in the limit $z\to 1$. In this case, neglecting the function $F_{-1}$, only the diagram A contributes so that we have
\begin{align}
\label{eq:resz1}
D_{q \to \psi} (z) \Big|_{z\to 1}&\simeq -4 \frac{\alpha_s^2 C_F}{M^3 z} \lb \frac{1}{\epsilon}-2-\ln\frac{\bar{z}M^2 }{\mu^2}\rb \frac{\langle \mathcal{O}^{\psi}(\Q{3}{S}{1}{8}) \rangle}{(d-1)(N_c^2 - 1)}
\,,
\end{align}
which agrees with the collinear integrated  fragmentation function in the same  limit calculated in ref.~\cite{Bodwin:2014bia}.

For completeness we have performed the same calculation in  momentum space. Here we only give the final result in $d=4$ dimensions,
\begin{equation}
\label{eq:TMD-FF-mom}
D_{q\to \psi} (z,\bmat{k}_{T};\mu,\zeta) = 2 \frac{\alpha_s^2 \;C_F}{\pi} \frac{\bar{z}}{M^3 z^3} \lb \frac{2 z^2}{\bar{z}} \frac{\bmat{k}_{T}^2}{(\bmat{k}_{T}^2 +  \bar{z} M^2 /z^2)^2}  + \frac{4}{\bmat{k}_{T}^2 + \bar{z} M^2 /z^2} \rb \frac{\langle \mathcal{O}^{\psi}(^3S_1^{(8)}) \rangle}{3 (N_c^2 - 1)}
\,,
\end{equation}
which can also be computed from the Fourier transform of the impact parameter space result in eq.~(\ref{eq:TMD-FF-b}),
\begin{equation}
D_{q\to H} (z,\bmat{k}_{T};\mu,\zeta) = \int \frac{d^2\bmat{b}}{(2\pi)^2} e^{-i\bmat{b} \cdot \bmat{k}_T}D_{q\to H} (z, b;\mu,\zeta)\,.
\end{equation}
This result holds formally when considering the explicit dependence on $\bmat{b}$ (i.e. with the strong coupling fixed), to perform the Fourier (or inverse-Fourier) transform. 
The integrand has an asymptotic $ \sim b \,J_0(bq)\,\exp(- c(M,z) b)$ behaviour at large $b$ and, thus, is convergent for $c(M,z) > 0$. Note that $c(M,z) =  M \sqrt{1-z}/z$ and thus the integral is convergent except the for cases $z\to 1$ and $M \to 0$. This is also apparent from the momentum space result since the $1/\bmat{k}^2_T$ divergence is regulated by the same coefficient.
We have confirmed this result by performing the integral numerically.

We will use the momentum space result in eq.~(\ref{eq:TMD-FF-mom}) in the following section in order to make the comparison of the quark-fragmentation and the photon-gluon fusion processes in the large $Q^2$ regime, $Q^2 \gg M^2\sim \bmat{q}_{T}^2$.

\begin{figure}[h!]
\includegraphics[width=\textwidth]{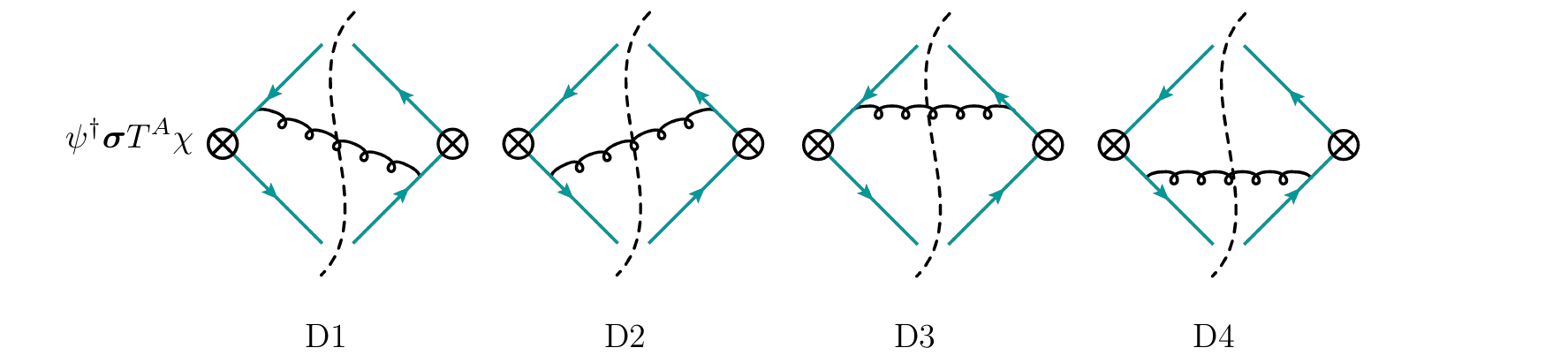}
\caption{\label{fig:fig4} The NLO diagrams for the $\Q{3}{S}{1}{8}$ LDME associated with the mixing through renormalization group evolution with the $P$-wave matrix elements.}  
\end{figure}

\subsection{Renormalization of LDMEs and channel mixing}
\label{sec:RGE-mix}

In this section we aim to rewrite eq.~(\ref{eq: nrqcd-match}) considering  resummation of logarithms of the ratio $\mu/\mu_f$, where $\mu_f$ is the scale at which the LDMEs are extracted. 
The most common choice for phenomenological extractions is $\mu_f = M$ and $\mu \sim 1/b$.  
In this study we consider $q_T \sim M$ and since $b$ is simply the conjugate variable of $q_T$ we should expect that $1/b \sim q_T$. 
Indeed the bulk in the impact-parameter-space distribution is found around this value, though significant support exists in the region $1/b \sim 0.5-20$~GeV (or equivalently $b\sim 0.05-2~\text{GeV}^{-1}$).\footnote{The exact range over which the distribution has support depends on the values of $q_T$ and $z$ we are looking.} 
Since we integrate over all values of the impact parameter, we find it important to perform the resummation of these logarithms.

A complication to this attempt comes from the mixing of the $\Q{3}{S}{1}{8}$ and $\Q{3}{P}{J}{8}$ mechanisms under renormalization group equations (RGEs). 
At fixed order the mixing occurs when a real ultra-soft gluon is exchanged between  heavy quarks (antiquarks) in the $\Q{3}{S}{1}{8}$ LDME at NLO. 
The emission of this ultra-soft gluon generates an ultraviolet (UV) divergence proportional to the LO $\Q{3}{P}{J}{1,8}$ LDMEs (see discussions in refs.~\cite{Petrelli:1997ge, Fleming:2019pzj}). 
Specifically we have that the order $\mathcal{\alpha}_s$ contributions to the  $\Q{3}{S}{1}{8}$  LDME is scaleless and proportional to the $P$-wave LO matrix element.  
Using the $1/\epsilon_{\UV} -1/\epsilon _{\IR}$ prescription we obtain for the sum of diagrams in figure~\ref{fig:fig4}:
\begin{equation}
\text{D1+D2+D3+D4} = \frac{4\alpha_s}{3 \pi m_c^2} \lbc C_F \sum_J \langle \Q{3}{P}{J}{1}  \rangle_{\text{LO}} +  B_F \sum_J \langle \Q{3}{P}{J}{8}  \rangle_{\text{LO}}  \rbc \; \lp \frac{1}{\epsilon_{\UV}} -  \frac{1}{\epsilon_{\IR}}\rp 
\,,
\end{equation}
where $B_F = (N_c^2 -4)/(4N_c)$ and the $\otimes$ vertex in figure~\ref{fig:fig4} corresponds to the bilinear operator: $\psi^{\dag} \bmat{\sigma} T^A \chi$. 
This UV divergence is removed through the renormalization of the LDMEs, which then leads to the following RGE:
\begin{equation}
\label{eq:LDME-rge}
\frac{d}{d \ln\mu} \langle \mathcal{O}_{\psi}(n) \rangle^{(\mu)} =  \sum_m \gamma^{nm}_{\mathcal{O}} \langle \mathcal{O}_{\psi}(m) \rangle^{(\mu)}\;.
\end{equation}
From eq.~(\ref{eq: nrqcd-match}) this immediately implies that the matching coefficients $d_{q\to Q\bar{Q}(n)}$ must also satisfy a non-diagonal RGE. A similar observation was made for the quarkonium TMD shape functions in ref.~\cite{Fleming:2019pzj}. The major difference here is that the non-trivial evolution is performed within the TMD fragmentation function and therefore the form of the factorized cross section is not modified from the standard case, in contrast to what happens in~\cite{Fleming:2019pzj}.
Therefore the RGE satisfied by the short-distance matching coefficients can be deduced from the consistency of the NRQCD factorization conjecture: 
\begin{equation}
\frac{d}{d \ln\mu}  d_{q\to Q\bar{Q}(n)} (z,b;\mu,\zeta) = \sum_m \lp \gamma_D \delta^{nm} + \gamma_{d}^{nm}) \rp  d_{q\to Q\bar{Q}(m)} (z,b;\mu,\zeta) 
\;,
\end{equation} 
where $\gamma_d = -\gamma_{\mathcal{O}}^T$.  
To test this equation we need to calculate the contributions to the quarkonium TMDFF at NLO, which is beyond the scope of this study. 
However, assuming that NRQCD factorization holds we may resum logarithms of the form $\ln (\mu/\mu_f)$ by solving eq.~(\ref{eq:LDME-rge}).
Following the notation of~\cite{Fleming:2019pzj} and up to next-to-leading-logarithmic (NLL) accuracy we have 
\begin{equation}
\label{eq:TMD-res}
D_{q \to \psi} (z,b;\mu,\zeta) = \frac{ d_{q \to Q\bar{Q}(\Q{3}{S}{1}{8})} (z,b;\mu,\zeta) } { N_{\text{col.}} N_{\text{pol.}} } 
\lb  \langle \mathcal{O}^{H} (\Q{3}{S}{1}{8}) \rangle^{(\mu_f)} -   \frac{24 B_F}{  \beta_0} \ln \lp \frac{\alpha_s(\mu)}{\alpha_s(\mu_f)}\rp  \frac{\langle \mathcal{O}^{H} (\Q{3}{P}{0}{8}) \rangle^{(\mu_f)}}{m_{Q}^2}  \rb
\;,
\end{equation}
where we have summed over $J=0,1,2$  and used $\langle \mathcal{O}^{\psi} (\Q{3}{P}{J}{8}) \rangle = (2J+1) \langle \mathcal{O}^{\psi} (\Q{3}{P}{0}{8}) \rangle$. The difference between eqs.(\ref{eq:NRQCD}) and (\ref{eq:TMD-res}) is the resummation of $\ln (\mu/\mu_f)$ and in the latter it is made explicit that TMD evolution is satisfied by the fragmentation function $D_{q\to{H}}$ and not the short distance matching coefficients $d_{q \to Q\bar{Q}}$. 

\begin{figure}
\centering
\includegraphics[width= \textwidth]{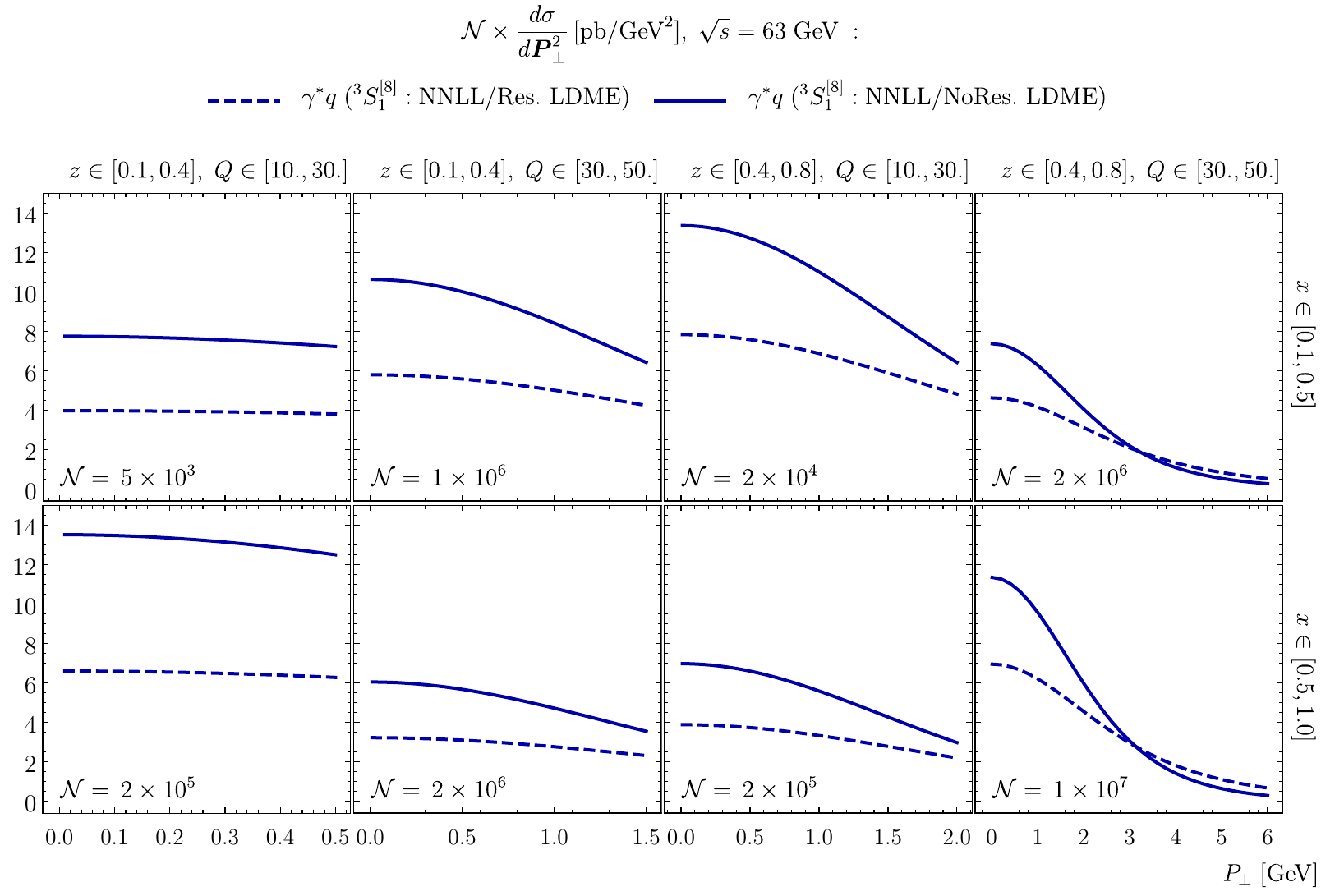}
\caption{\label{fig:NoResvsRes_63} The effect of the NLL resummation of $\ln (m_Q b)$ in the  quarkonium cross section coming from TMD fragmentation ($\gamma^* q$)  for EIC kinematical settings $\sqrt{s}=63$ GeV.}
\end{figure}

We demonstrate the effect of the resummation of $\mu/\mu_f$ logarithms in figure \ref{fig:NoResvsRes_63}, where we show the cross-section for the EIC kinematics at $\sqrt{s} = 63$~GeV, using the fixed LDMEs versus their evolved values. 
We use the BCKL set of LDMEs as shown in table~\ref{tb:ldme}. 
It is interesting to note that, as expected, the effect of resummation decreases in the region $P_\perp \sim M$.

\subsection{On quarkonium  TMDFFs from a gluon}
\label{sec:gluon-TMD}

This study focuses primarily on quarkonium fragmentation in SIDIS and, for this process, the main interest is the light-quark fragmentation functions. 
However, a brief mention to the gluon fragmentation process is in order, since this formalism can also be applied in phenomenologically interesting processes beyond DIS. 
Particularly, gluon fragmentation will play an important role in quarkonium fragmentation within jets, which has attracted significant interest in recent years. 
This process can be accessed in hadron colliders and particularly at the LHC, where jets are produced in abundance. 
Measurements~\cite{Aaij:2017fak, Diab:2019she} of the momentum fraction $z_{T}\equiv p_T^{H}/p_T^\jet$ for in-jet quarkonia have already revealed interesting aspects of quarkonium production and a striking deviation from Monte-Carlo implementations.
An explanation for these deviations was given in ref.~\cite{Bain:2017wvk} and a better qualitative agreement with the data was provided through the fragmentation picture of quarkonium production at large transverse momenta. 
A multi-differential aspect of these measurements, such as including the transverse momentum w.r.t. the jet axis, will enable a three-dimensional picture of quarkonium fragmentation and an in-depth  understanding of its production mechanisms. 
In the massless limit, i.e. $q_T \gg M$, the quarkonium TMD fragmentation within jets was studied in refs.~\cite{Bain:2016rrv, Makris:2017hjk}. The framework discussed here can be extended  and applied to such studies in the small transverse momentum limit, $q_T \lesssim M $.  

The LO contribution to gluon TMDFFs comes from the color-octet $\Q{3}{S}{1}{8}$ channel through a simple decay of an off-shell gluon to the heavy quark-antiquark pair.  
At NLO, contributions from both $\Q{1}{S}{0}{8}$ and $\Q{3}{P}{J}{8}$ channels are introduced.  
The color-singlet channel $\Q{3}{S}{1}{1}$ appears only at NNLO. 
Although these contributions are only relevant at higher orders in the perturbative expansion, recent phenomenological studies~\cite{Bain:2017wvk}  of in-jet quarkonia suggest that numerical enhancements from the LDMEs require all four channels to describe the spectrum of quarkonia within jets. 

Furthermore, for the $\Q{3}{S}{1}{8}$ channel, light-quark fragmentation and gluon fragmentation are inevitably related, since at this order the heavy quark-antiquark pair can only be produced by a subsequent decay of an off-shell gluon (see figure~\ref{fig:B_fig2}). 
It is therefore implied that the channel mixing, through RG evolution as discussed in section~\ref{sec:RGE-mix}, will also be relevant for gluon TMDFFs. 

\section{Quarkonium from photon-gluon fusion}
\label{sec:gluon-photon}

The effect of photon-gluon fusion in the differential cross section that we study is highly complex in the TMD framework that we are using, because higher-twist gluon operators appear already at lowest order. 
To organize the discussion we consider two separate regions of $z$: region-1 where  $(1-z) \ll 1$ and region-2 where $(1-z) \sim 1$. In region-1 one can study the small transverse momentum region $\bmat{q}_{T}^2 \ll M^2$ from channels $\Q{1}{S}{0}{8}$ and $\Q{3}{P}{0,2}{8}$ within the framework of shape-functions in SCET$_Q$ \cite{Echevarria:2019ynx, Fleming:2019pzj}. 
In addition, sub-leading operators in SCET$_Q$ can give contributions from the channels $\Q{3}{S}{1}{1}$ and $\Q{3}{S}{1}{8}$. 
While these contributions are sub-leading in the $\bmat{q}_{T}^2 / Q^2$ power-counting, the color-singlet channel can get significant enhancement of order $1/v^4$, where $v$ is the relative velocity of the heavy quark-antiquark pair in the quarkonium rest frame. 
A proper description of region-1 is still in progress.  

In this work we focus on region-2. 
In this case beyond the quark fragmentation that is discussed in section~\ref{sec:quark}, we also have a contribution from the color-singlet channel from photon-gluon fusion process. 
In this section we give an estimate of this part of the cross section coming from perturbative QCD, that is, looking at the hard part of this process. 
In this way we get an order of magnitude understanding of both quark fragmentation and photon-gluon fusion (only color-singlet channel). 
The cross section that we are looking for is the one of the process depicted in figure~\ref{fig:B_fig3} (b), already studied a long ago in \cite{Merabet:1994sm}.

Considering the result for the cross section in \cite{Merabet:1994sm} integrated over all angles and in the limit $Q^2\gg M^2 \sim \bmat{q}_{T}^2$, and expressing the result with the help of \cite{Berger:1980rj}  one gets
\begin{align}
\label{eq:FO-app}
\frac{d\sigma (\gamma^* g)}{dx  dz dQ^2 d \bmat{P}_\perp^2} \simeq \frac{\alpha_s^2(\mu) \,\alpha^2_{\rm em} \pi}  {(1-\varepsilon) s^2}  \frac{64 \;e_H^2 }{27 z \bar z^3(2-z)^2} 
\frac{1}{ x^2}\frac{\langle \mathcal{O}^{\psi}(\Q{3}{S}{1}{1}) \rangle}{ Q^2 M^3 }  f_{g \leftarrow h} (x;\mu^2)  F_g(\bar z, \frac{P_\perp }{M})
\,,
\end{align}
where
\begin{align}
F_g(a,b)=\frac{a^4+a^2+b^2}{(1+b^2/a^2)^2}
\end{align}
and $e_H=2/3$ ($e_H=1/3$) for charmonium (bottomonium).

We would like to compare this result with the quark fragmentation process in the same limit ($Q^2\gg M^2 \sim \bmat{q}_{T}^2$).
The limit of  $Q \gg M\sim q_T$ 
at leading power in the $\sim q_T/Q$ or $M/Q$ expansion, is obtained by the TMD factorization formula in eq.~(\ref{SIDIS:xSec}) convolving all terms at their fixed-order components. 
At LO this result is rather simple since the hard coefficient is just 1 and the TMDPDF is proportional to the collinear PDF $\sim f(x,\mu) \delta^{(2)}(\bmat{q}_T)$. 
Thus, convolving with the LO TMDFF from eq.~(\ref{eq:TMD-FF-mom}) we have
\begin{equation}
\frac{d\sigma(\gamma^* q)}{dx  dz dQ^2 d \bmat{P}_\perp^2} \simeq \frac{\alpha_{\text{em}}^2\, \pi }{(1-\varepsilon) s^2} \sum_f\frac{2\pi e_f^2 }{x^2 }  f_{f \leftarrow h} (x;\mu^2) D_{f \to H} (z, \bmat{q}_{T} ;\mu).
\end{equation}
With further reorganization of this result to match the form of eq.~(\ref{eq:FO-app}) we find
\begin{equation}
\label{eq:TMD-app}
\frac{d\sigma(\gamma^* q)}{dx  dz dQ^2 d \bmat{P}_\perp^2} \simeq \frac{\alpha_s^2 (\mu)\,\alpha_{\text{em}}^2\, \pi }{(1-\varepsilon) s^2} \sum_f \frac{4 \,e_f^2}{9 x^2\, z } \frac{\langle \mathcal{O}^{\psi}(\Q{3}{S}{1}{8}) \rangle }{M^5} f_{f \leftarrow h} (x;\mu^2) F_q (z,\frac{\bmat{P}^2_{\perp} }{M^2 \bar{z}})
\,,
\end{equation} 
where 
\begin{equation}
F_q(a,b) = \frac{a^2 b+2(1-a)(1+b)}{(1-a)(1+b)^2}
\,.
\end{equation}
This equation allows us to have an order of magnitude comparison of these two contributions to the cross-section:
\begin{equation}
\label{eq:proc-scaling}
\frac{d\sigma(\gamma^* g)}{d\sigma (\gamma^* q)} \sim  \lp \frac{M}{Q v^2} \rp^2
\,,
\end{equation} 
where for the relative velocity $v$ we have $v^2 \sim 0.3$ for charmonium states and $v^2 \sim 0.1$ for bottomonium. 
In table~\ref{tb:rel-contr} we give the values of this estimate for different values of $Q$. 
We find that for charmonium states and for $Q>20$ GeV the photon-gluon fusion (only color-singlet channel) contributes less than 10\% compared to the quark fragmentation process. 
The numerical relative contributions of the full fixed-order photon-gluon fusion in color-singlet channel and the light-quark fragmentation processes, which are sensitive to the values of the LDMEs and collinear PDFs, are discussed in the following section.
 
\begin{table}[t!]
\renewcommand{\arraystretch}{1.7}
\begin{center}
\begin{tabular}{|l|c|c|c|c|}
\hline
$ M^2/ (Q^2 v^4 ) $ & $Q=10\;[\text{GeV}]$ & $Q=30\;[\text{GeV}]$  & $Q=50\;[\text{GeV}]$& $Q=100\;[\text{GeV}]$  \\ \hline \hline
charmonium ($v^2 \sim 0.3$)   &  1.0  & 0.1  &  0.04 &0.01 \\ \hline
bottomonium ($v^2 \sim 0.1$)   &  n.a.  & 10.0  & 3.8&1.0  \\ \hline
\end{tabular}
\caption{  \label{tb:rel-contr} Order of magnitude estimate of the relative contributions from the quark fragmentation and photon-gluon fusion processes.}
 \end{center} 
\end{table}
 
\section{Numerics}
\label{sec:numerics}

The relative importance of photon-gluon fusion and quark fragmentation for different intervals of the kinematic variables is highly non-trivial. 
In order to plot the contribution to the cross section for the quark fragmentation we need some knowledge of the TMDPDFs as provided by other processes. 
We use here the $\zeta$-prescription \cite{Scimemi:2018xaf} and the phenomenological results of ref.~\cite{Scimemi:2019cmh}. 
The $\zeta$-prescription allows to express the cross section in terms of scale-independent quantities: in this way we implement the TMD evolution kernel at N$^3$LL,  the TMDPDFs at NNLO, and the quarkonium TMDFF at LO, which we have calculated in this work. 
The non-perturbative parameters for the TMDPDF are chosen consistently with the PDF set NNPDF31\_nnlo\_as\_0118~\cite{Ball:2017nwa}. 
There is not yet a consensus on the correct values of LDMEs and even though they are assumed to be universal, extractions from different experiments seem to contradict this assumption. 
In this sense the present analysis can help also to give more information on this issue. 
In order to show the impact of the LDME set, we use three different choices. 
As a first, we use extractions from a large $p_T^\psi$ analysis in hadronic collision~\cite{Bodwin:2014gia}. 
In this analysis the authors used the NRQCD fragmentation functions to describe the quarkonium $p_T$ spectrum. 
Then, we also include a comparison from a global fit analysis~\cite{Butenschoen:2011yh} and a large $p_T$ FO NRQCD analysis from ref.~\cite{Chao:2012iv}. The numerical values for the LDMEs that we use are given here in table~\ref{tb:ldme}. We evolve the $S$-wave color-octet values according to eq.~\eqref{eq:TMD-res}.

\begin{table}[h!]
\renewcommand{\arraystretch}{1.5}
\begin{center}
\begin{tabular}{|l|c|c|c|}
\hline
& $\langle\mathcal{O}^{J/\psi}(^3S_1^{[1]}) \rangle$  & $\langle \mathcal{O}^{J/\psi}(^3S_1^{[8]}) \rangle $ & $\langle \mathcal{O}^{J/\psi}(^3P_0^{[8]}) \rangle/m_c^2 $  \\
\hline
BCKL \cite{Bodwin:2014gia}      & $1.32\pm 0.20$ & $(1.1\pm 1.0) \times 10^{-2}$    & $(0.49\pm 0.44) \times 10^{-2}$  \\ 
\hline
B\&K \cite{Butenschoen:2011yh}  & $1.32\pm 0.20$ & $(0.224\pm 0.59) \times 10^{-2}$ & $(-0.72\pm 0.88) \times 10^{-2}$  \\
\hline
CMSW \cite{Chao:2012iv}         & $1.32\pm 0.20$ & $(0.30\pm 0.12) \times 10^{-2}$  & $(0.56\pm 0.21) \times 10^{-2}$  \\
\hline
\end{tabular}
\caption{\label{tb:ldme}  
LDMEs for NRQCD production mechanisms we use in this paper in units of ${\rm GeV^3}$.  Note that we use the same color-singlet matrix element even for the CMSW extraction where a slightly smaller value (1.16) was used. This will not influence our conclusion regarding the relative importance of each channel. }
\end{center}
\end{table}

We have considered two possible settings for the EIC, at $\sqrt{s}=63$~GeV and $\sqrt{s}=140$~GeV, which are expected to be typical for this collider~\cite{Accardi:2012qut}.
In both settings we calculate the contribution to the cross section given respectively by photon-gluon fusion to color singlet and quark fragmentation processes. 

The relative contribution of each input depends crucially on the values of $Q$, $x$ and $z$. Using the BCKL set of LDMEs we find that the general trend is: the quark-fragmentation dominates for high $Q$, high $x$ and low $z$. The enhancement at large $Q$ is expected from the scaling relation in eq.~(\ref{eq:proc-scaling}), while the enhancement at large $x$ is purely due to the values of the PDFs (gluon PDF in photon-gluon fusion is larger that quark PDFs in quark fragmentation at low $x$, and vice-versa).
For the CMSW LDMEs the color-octet is suppressed by a factor of four and thus the importance of this channel is reduced. Particularly, we find no kinematic region where the octet-dominates. However, for most regions we find that its contribution is comparable to the singlet and thus necessary to include for a consistent analysis. On the other hand, the B\&K LDMEs yield a rather suppressed contribution in most regions. 
Also note that for $P_\perp \gtrsim M$ it yields a negative cross-section, due to the negative contribution from the $P$-wave LDME. 
It is unclear whether this will change when including power-corrections from the fixed-order (NRQCD) cross-section.

\begin{figure}
\centering
\includegraphics[width= \textwidth]{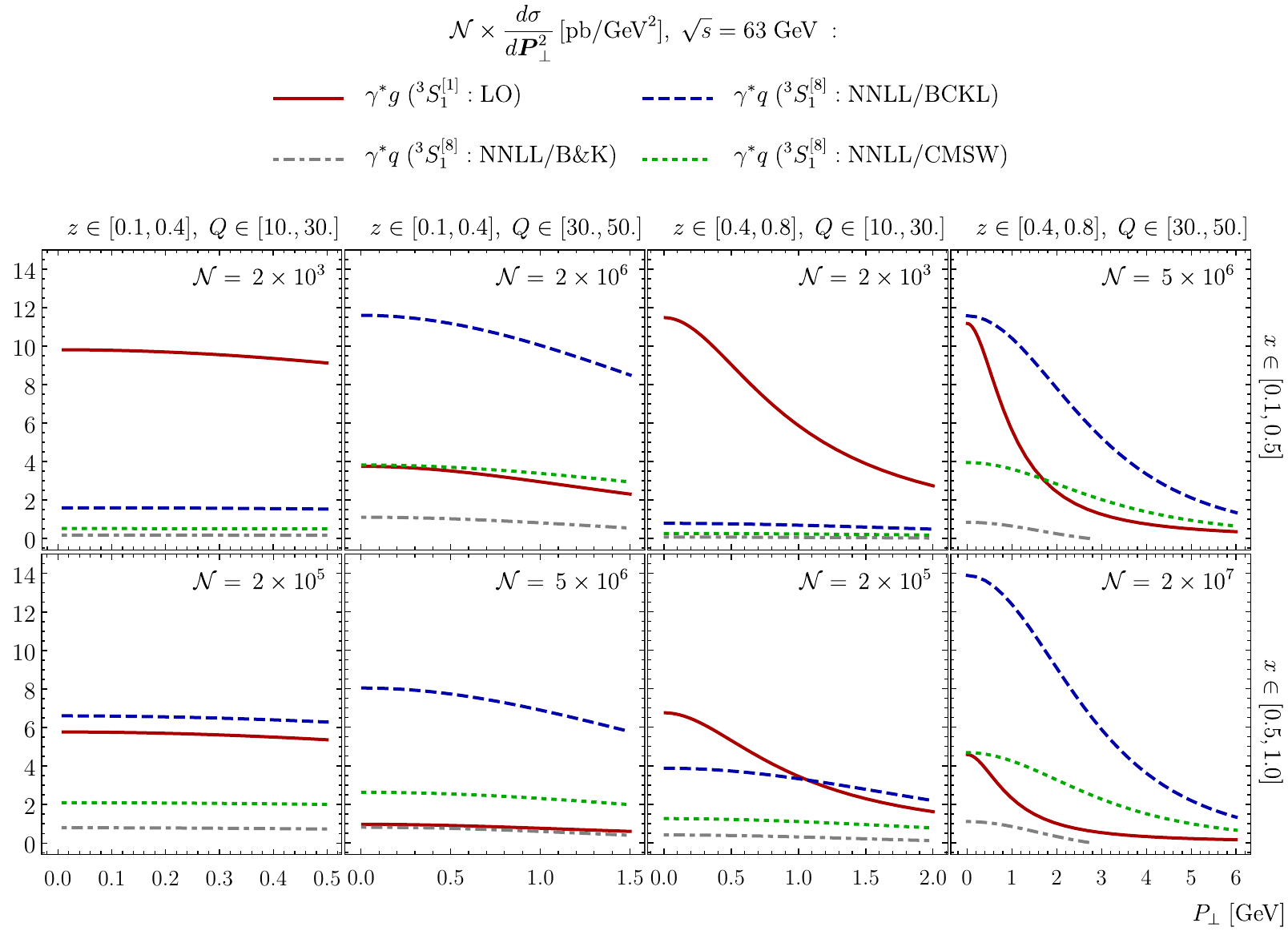}
\caption{\label{fig:FOvsTMD_63} The quarkonium cross section coming from photon-gluon fusion in the color-singlet channel at leading order  ($\gamma^* g$) and TMD fragmentation ($\gamma^* q$)  for EIC kinematical settings $\sqrt{s}=63$ GeV.}
\end{figure}

\begin{figure}
\centering
\includegraphics[width= \textwidth]{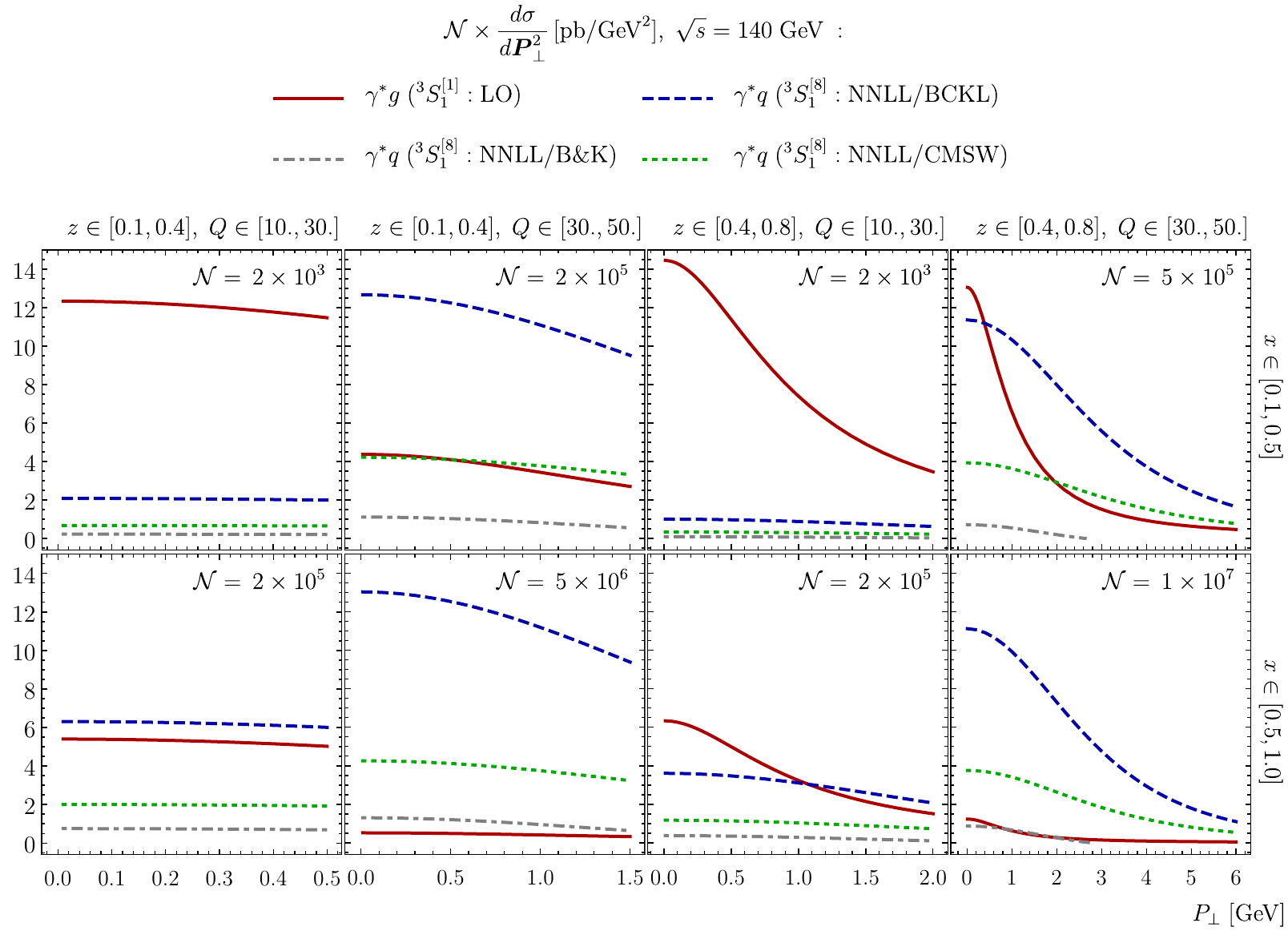}
\caption{\label{fig:FOvsTMD_140} The quarkonium cross section coming from photon-gluon fusion in the color-singlet channel at leading order ($\gamma^* g$) and TMD fragmentation ($\gamma^* q$)  for EIC kinematical settings $\sqrt{s}=140$ GeV.}
\end{figure}

We have plotted  the cross section integrated over several intervals of these variables to show this effect more quantitatively.
The results are shown in figures~\ref{fig:FOvsTMD_63} and~\ref{fig:FOvsTMD_140}.
For each of the variables, $z$, $x$ and $Q$ we chose two possible bins as shown in the figures, resulting in eight distributions for each center-of-mass energy.  
For all kinematic regions, in order to stay within the TMD-regime, we plot the transverse momentum distributions only in the region $P_{\perp} \in [0, z_{\text{min}} Q_{\text{min}}/2]$, where $z_{\text{min}}$ and $Q_{\text{min}}$ are the lower bounds of the corresponding bins.

\section{Conclusion}
\label{sec:outlook}

Quarkonium production at the Electron-Ion Collider is generally considered as one of the most important processes for the study of gluon TMDs. 
We have shown that the physics for these heavy-quark events is highly non-trivial.

In this paper we have studied the quarkonium transverse-momentum-dependent fragmentation functions (TMDFFs) within the framework of non-relativistic QCD (NRQCD).  
We decompose the TMDFFs in a similar way as done in the corresponding collinear fragmentation functions~\cite{Braaten:1996rp}. 
We use the NRQCD factorization conjecture to write TMDFFs as a sum of products of short-distance matching coefficients and the standard NRQCD long-distance matrix elements (LDMEs).  

Focusing on the case of $J/\psi$ production in semi-inclusive deep-inelastic scattering (SIDIS) and for $Q$ parametrically larger than the mass of the heavy quark, we use the standard TMD factorization to describe the cross section as a convolution of TMD parton distribution functions and the quarkonium TMDFFs. 
We provide the leading-order term in the perturbative expansion of the missing quarkonium TMDFF and use this result to make numerical comparisons against the competing photon-gluon fusion channel (through color-singlet intermediate $Q\bar{Q}$ state). 
For this process, only the light-quark fragmentation functions are relevant, and at LO we only have a contribution from the $\Q{3}{S}{1}{8}$ channel. 
We also included a short discussion on channel mixing through renormalization group evolution and phenomenological applications of the gluon TMDFFs at the LHC. 

For our numerical implementation of the fragmentation cross section we use the public Artemide code~\cite{web}, which we modified appropriately to incorporate the new quarkonium TMDFF.
We discussed the kinematic regions of the future EIC and we provided two panels of plots for $\sqrt{s}=63$~GeV and $\sqrt{s}=140$~GeV. 
We showed that photon-gluon fusion is usually dominant at low values of $Q$ and $x$, while for some choices of LDMEs quark fragmentation plays a major part in the opposite limit. 
The theoretical precision is at the moment limited only by the final quarkonium fragmentation function. 
In the case of (color-singlet) photon-gluon fusion instead, a factorization theorem has not been yet formulated and its feasibility should be investigated in the future.

\section*{Acknowledgements}
We thank A. Vladimirov for discussions on Artemide code.
M.G.E. and I.S. are supported by the Spanish Ministry grant PID2019-106080GB-C21. 
This project has received funding from the European Union Horizon 2020 research and innovation program under grant agreement Num. 824093 (STRONG-2020). 
Y.M. is supported by the European Union’s Horizon 2020 research and innovation program under the Marie Sk\l{}odowska-Curie grant agreement No. 754496-FELLINI.

\bibliographystyle{JHEP}
\normalbaselines 
\bibliography{TMD_ref}

\end{document}